# ODOO-BASED SUBCONTRACT INTER-SITE ACCESS CONTROL MECHANISM FOR CONSTRUCTION PROJECTS
# (Cơ chế quản lý truy cập giữa các công trình của nhà thầu phụ trên nền Odoo)

Huy Hùng Hồ [1,2,3,*]_2352421_Computer Science, Nhân Lê Thành [1,3]_2011738_Computer Science,
Nam Nguyen Hong [1,3], Phuong-D Nguyen[1,3]

[1] Faculty of Computer Science and Engineering, Ho Chi Minh City University of Technology (HCMUT), 268 Ly Thuong Kiet Street, Dien Hong Ward, Ho Chi Minh City, Vietnam

[2] Office for International Study Programs, Ho Chi Minh City University of Technology (HCMUT), 268 Ly Thuong Kiet Street, Dien Hong Ward, Ho Chi Minh City, Vietnam

[3] Vietnam National University Ho Chi Minh City, Tan Lap Quarter, Dong Hoa Ward, Ho Chi Minh City, Vietnam

* Corresponding author: hung.hokhmtclc@hcmut.edu.vn

## Abstract

In the era of Construction 4.0, the industry is embracing a new paradigm of labor elasticity, driven by smart and flexible outsourcing and subcontracting strategies. The increased reliance on specialized subcontractors enables companies to scale labor dynamically based on project demands. This adaptable workforce model presents challenges in managing hierarchical integration and coordinating inter-site collaboration.

Our design introduces a subsystem integrated into the Odoo ERP framework, employing a modular architecture to streamline labor management, task tracking, and approval workflows. The system adopts a three-pronged approach to ensure synchronized data exchange between general contractors and subcontractors, while maintaining both security and operational independence. The system features hybrid access control, third-party integration for cross-domain communication, and role-based mapping algorithm across sites. The system supports varying degrees of customization through a unified and consolidated attribute mapping center. This center leverages a tree-like index structure and Lagrange interpolation method to enhance the efficiency of role mapping.

Demonstrations highlight practical application in outsourcing, integration, and scalability scenarios, confirming the system's robustness under high user volumes and in offline conditions. Experimental results further show improvements in database performance and workflow adaptability to support a scalable, enterprise-level solution that aligns with the evolving demands of smart construction management.

## 1. Introduction

As globalization advances, data and services are no longer confined to isolated domains. It is driving the construction industry toward the next generation construction 4.0 to encourage subcontractor outsourcing. The collaboration context led by government agencies and their organizations is employed in developing smart city infrastructures. These infrastructures aim to leverage intelligent communication infrastructure and cloud service providers to perform heavy tasks in distributed outsourcing in outsourcing domain environments. In such a distributed domain manner, users, devices, and services spanning across distinct administrative boundaries benefit from advantages such as elasticity, resource sharing, and operational efficiency. It also emerges as a critical challenge of security in access control across different domains which requires **inter-site access control** mechanisms. Due to its native security theme, we use the terms cross-domain and inter-site interchangeably.

Cross domain allows users from a domain to access resources that are stored in a different domain with its own trusted relationships, mapped attributes, and access policies while preserving the autonomy and security requirements of each domain [1]. These collaborations would require duplicative identity management systems which result in the management costs of redundancy and loose coupling separation [2]. We are interested in developing an efficient and secure cross-domain access control mechanism which allows integrating multi collaborating sites in a secure digital ecosystem [3]. The main contributions of this work are:

- Design and implement an Odoo-based inter-site access control management subsystem.
- Propose a cross-domain access control mapping scheme based on proxy re-encryption algorithm.

## 2 Inter-site access control management

In this section, we introduce the design of on-flying task management and then we provide an implementation of an access control management subsystem.

### 2.1. On flying task management

We aim to design for enterprise scale with emphasis on performance and load capacity by leveraging inter-access and task delegation among sites. This module supports the adaptive workflow [4] where project personnel can collaborate and communicate tasks through a construction approval process. It supports:

- **Task outsourcing**: a task can be processed by an outsourcing partner which implies the assignment to a subcontractor.
- **Subcontractor delegation**: a matching subcontractor can be chosen via external negotiations to create connection to the partner subcontractor using the cross-domain mapping.
- **Inter-site task tracking mechanism**: map and validate tracked cross domain spanning tasks.

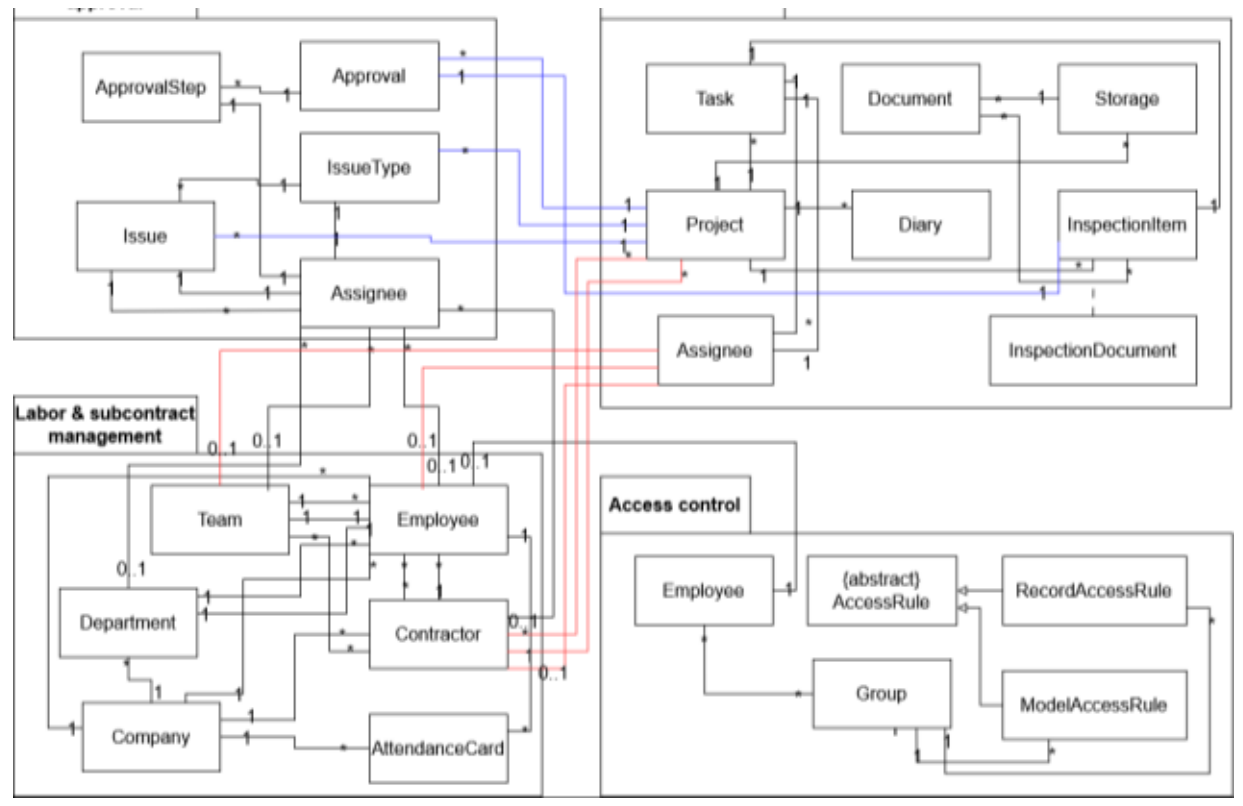

***Figure 1.*** *Access control and tracking subsystem*

### 2.2. Access control management subsystem

Each site includes an access control and racking subsystem as in figure 1. The subsystem allows both the subcontractor and the general contractor to manage their labor. The general contractor can delegate the task to different subcontractors. Therefore, its design integrates both sites: contractor site and subcontractor site. Based on the existing and separated site, the subcontractor site can operate independently from the contractor site.

In the approval process, the initiator will first create a request including a deadline. The approver performs the review and matching until the approval process is complete. In the issue tracking process, issues are initially created by the initiator, which is not yet visible to other personnel. These personnel issues are assigned by resolvers through a matching process. Finally, the resolution result is verified and the issue is closed.

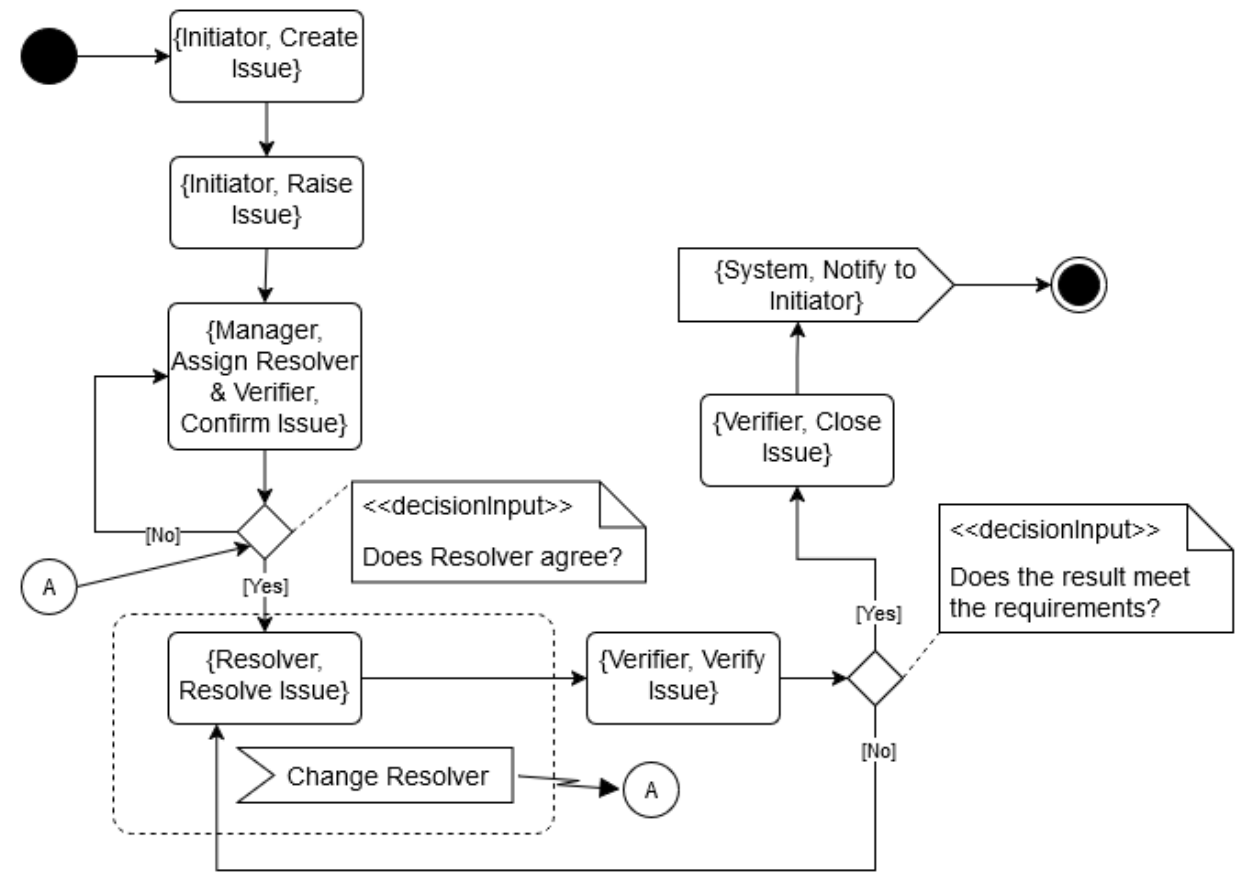

***Figure 2.*** *Issue tracking activity diagram*

## 3. Proposed access control mapping mechanism

### 3.1. Overview of Proxy Re-Encryption (PRE)

Our scheme inherits the PRE scheme in [1], and we combine some of the components upgrading in [2] to the PRE algorithm. It enhances access control, including the following steps:

(1) $Setup(\lambda) \rightarrow (pk, sk)$ to generate a tuple of system private keys - secret keys $(pk, sk)$.

(2) $KeyGen(A, sk) \rightarrow SK_A$ to generate the user's private key of attributes A.

(3) $Encrypt(pk, (M, \rho), K_{sym}) \rightarrow CT$: Data Owner (DO) encrypts the message to create ciphertext CT..

(4) $ReKeyGen(SK_A, PK_{A'}) \rightarrow rk_{A\rightarrow A'}$: a Data Requester (DR) from a different domain executes ReKeyGen() algorithm using its public key $PK_{A'}$ to re-encryption key $rk_{A\rightarrow A'} = \{rk_1, rk_2, rk_3\}$

(5) *ReEncrypt(PK*$PK_{A'}$*, *$rk_{A\rightarrow A'}$*, CT, (M', ρ')) → CT'*: DO sends the re-encryption key to the proxy to generate the re-encrypted ciphertext *CT'*.

(6) *Decrypt(*$PK_A$*, *$SK_A$*, CT) →* $K_{sym}$: proxy takes $SK_2$, $SK_3$ and $CT$ to reconstruct the symmetric key $K_{sym}$ using bilinear mapping operations.

## 3.2. Proposed access control mapping scheme

**A. *Setup(λ) → (sk, pk)*:**

This step mainly follows the derivative keys scheme based on the security parameter $\lambda$ to generate a couple of master secret keys $sk$ and the public key $pk$ for each session indexed by k in the next step.

- Given two multiplicative cyclic groups of prime order $p$, namely $G_0$. Randomly choose a prime number $p$ to establish a finite field $Z^*_p$. Define a bilinear mapping: $e = G_0 \times G_0 \rightarrow G_T$.
- Define a generator $g$ of these cyclic groups.
- Security parameter $\lambda = \{A, b\}$ consists of the set of DO's attributes $A$, and a protection attribute $b$ added for securing outsourced computations.
- Create a master secret key $sk$:
  + Proxy randomly picks $m, n \in Z^*_p$. The set $sk = \{m, n\}$ forms the master secret key. $m$ is used to blind the message key in the pairing $e(g, g)^m$, while $n$ is useful for creating partial keys for users.
  + This master $sk$ is kept secret by the proxy.
- Construct a public key $pk$:
  + Pass the cyclic group $G_0$ and its generator $g$.
  + Compute $h = g^n$, which is used in key blinding and encryption. DO and DR can use it.
  + Compute $e(g, g)^m$, which is a crucial part of the encryption. Only users with the right key structure can recover $e(g, g)^{mv}$ to decrypt and get the original ciphertext CT.
  + For $a_i$ in $A$: select a random number $p_i \in Z^*_p$ set $W_i = g^{p_i}$, as the public key associated with $a_i$.
  + Protection attribute $b$: select $p_b \in Z^*_p$ set $W_b = g^{p_b}$.
  + Send the obtained public key $pk = \{G_0, g, h, e(g, g)^m, \{W_i = g^{p_i}\}_{i=1}^{len(A)}, W_b\}$ to DO.
  + Construct the pair $(sk, pk)$.

***Remark 3.2.1.1*** **(infeasible reconstruction)** the procedure described in section ***3.2.A.Setup(λ)*** ensures outsourced servers cannot reconstruct the full secret without the proper leaf node being satisfied.

**B. *KeyGen(A, sk) →*** $SK_A$**:**

In this step, we base on $sk$ to create session keys with the three tuples. To ensure the nonce property, we use one-time usage random number $k$ and combine with a nonce $v$ to reconstruct the session keys. Then, we construct a set of secret keys $SK_A = \{SK_1, SK_2, SK_3\}$.

- Construct the first tuple $SK_1$:
  + The proxy chooses a random number $k \in Z^*_p$ and get $m$ and $n$ from the master secret key $sk$ to compute $SK_1 = g^{\frac{m+k}{n}}$ that serves for the decryption phase afterwards.
- Construct the second tuple $SK_2$:
  + The proxy chooses a random number $kb \in Z^*_p$ that is associated with the protection attribute $b$. Get $W_b$ from the public key to calculate $SK_2 = \{g^{kb}, g^k \cdot W_b{}^{kb}\}$. This eliminates a user with only attribute keys but no protection attribute.
- Construct the last of the three tuples $SK_3$:
  + For $a_i$ in $A$, we randomly select $k_i \in Z^*_p$ associated with each attribute and get $SK_{3,i} = \{g^{k_i}, g^k \cdot W_i{}^{k_i}\}$. Then we aggregate them to create $SK_3$ which is the set of all $SK_{3,i}$: $SK_3 = \{SK_{3,i}\}_{i=1}^{len(A)}$. This helps prevent collusion caused by two users attempting to combine their partial keys since each $g^{k_i}$ is tied to a different random $k$. Besides, it ensures correct attribute possession by eliminating the reconstruction a secret using Lagrange interpolation which requires enough correct attributes and their corresponding keys.

Finally, we arrive at the secret key of attribute $A$: $SK_A = \{SK_1, SK_2, SK_3\}$.

**C. *Encrypt(pk, (M, ρ),*** $K_{sym}$***) → CT={*** $CT_1$***,*** $CT_2$***,*** $CT_3$***}***

In this step, DO's public key and access control (M, $\rho$) are combined with a symmetric key to encrypt messages. We define matrix M of size *r* x *c* with piecewise mapping to each attribute by using nonce *v* as blind parameter to establish $\vec{u} = (v, y_2, \ldots, y_c)$ in mapping function $\rho$: $v_j = M_j \cdot \vec{u}, j \in [1, r]$. More specific:

- DO selects $v \in Z^*_p$. Compute $CT_1 = \{A_1, A_2\}$, which is a set $\{K \cdot e(g, g)^{mv}, g^{nv}\}$. It acts as the key protection thanks to the usage of the blinding parameter $v$.
- Build a matrix of size *r* x *c* where:
  + *r*: number of leaf nodes + 1. An extra row is kept for the protection attribute *b*.
  + *c*: number of variables needed to represent the access policy tree as a linear system.
- Setup a vector $\vec{u} = (v, y_2, \ldots, y_c)$ where $y_2, \ldots, y_c$ are randomly chosen from $Z^*_p$. Map each row of $M$ to an attribute:
  + Suppose $i_b$ is the row that contains the protection attribute. Calculate $vb = M_{i_b} \cdot \vec{u} \rightarrow CT_2 = \{B_b, C_b\} = \{g^{vb}, W_b{}^{vb}\}$, which is the protection attribute ciphertext component.

+ For all rows $i \neq i_b$, calculate $v_i = M_i \cdot \vec{u} \rightarrow CT_3 = \{B_i, C_i\}_{i=1}^{r} = \{g^{v_i}, {W_i}^{v_i}\}_{i=1}^{r}$. This set is the shared attributes for leaf nodes.

Then, DO obtains the ciphertext $CT = \{CT_1, CT_2, CT_3\}$.

### D. *ReKeyGen($SK_A$, $PK_{A'}$) → $rk_{A\rightarrow A'}$*:

We will generate a new sequence of keys for our ciphertext chain using DO's secret key $SK_A$ and DR's public key $PK_{A'}$. The form of our sequence of re-encrypted keys: $rk_{A\rightarrow A'} = \{rk_1, rk_2, rk_3\}$.

- Map to $CT_1$: Randomly choose $\alpha \in Z^*_p$. $\beta'$ is DR's own key → $PK_{A'} = g^{\beta'}$. Calculate $rk_1 = g^{\alpha} \cdot g^{\beta'}$ .
- Map to $CT_2$: Randomly choose $\mu \in Z^*_p$. Calculate $rk_2 = g^{\alpha} \cdot {W_b}^{\mu}$ for the protection attribute.
- Map to $CT_3$: Randomly choose $t \in Z^*_p$. Use $k_i$ when generating $SK_{3,i}$, we can calculate $rk_3 = g^{k_i \cdot t}$ for each attribute in A.

### E. *ReEncrypt($PK_{A'}$, $rk_{A\rightarrow A'}$, CT, (M', ρ')) → CT'*

Since DR and DO are in different domains, their access controls are completely distinct. We thus need to re-encrypt our ciphertext to align with DR's access control ($M'$, $\rho'$). By doing so, DR can easily decrypt in the *Decrypt()* phase with Lagrange Interpolation. $M'$ and $\rho'$ are defined similarly to $M$ and $\rho$ during the *Encrypt()* phase; the only difference is their initialized value.

- The proxy randomly selects a new secret key to share, denoted as $v' \in Z^*_p$. Build an access matrix $M'$ of size $r'$ x $c'$ column defined in the same way as $M$. Further initialize the vector $\overrightarrow{u'} = (v', y'_2, \ldots, y'_{c'})$ where $y'_2, \ldots, y'_{c'}$ are randomly chosen from $Z^*_p$.
- We first re-encrypt $CT_1$ as $CT'_1 = \{A'_1, A'_2\}$ where:
  + $A'_1 = \frac{A_1 \cdot (g^{\beta'})^{s'} \cdot e(g^{\beta}, rk_3) \cdot e(A_2, rk_1)}{e(A_2, rk_2) \cdot e(A_2, g^{\beta'})}$, $\beta'$ is the DR's own key. and $A'_2 = g^{v'}$.
- Suppose $i_b$ is the row that contains the protection attribute. For all rows $i \neq i_b$, calculate $v'_i = M'_i \cdot \vec{u}'$ → $CT'_3 = \{B'_i, C'_i\}_{i=1}^{r'} = \{g^{v'_i} \cdot H(\rho'(i))^{-r'_i}, g^{r'_i}\}_{i=1}^{r'}$ where:
  + $r'_i$ is arbitrarily chosen for each attribute from $Z^*_p$.
  + $H(x$: str) is a cryptographic hash function that receives a string as input, and maps it to our multiplicative group since H: $\{0,1\}^* \rightarrow G_0$. With a negative exponent, it can resolve collisions.
  + $\rho'(i)$ returns a string of each row $i$ of matrix $M'$.
- Finally, for $CT'_2$, it is computed in the same way as $CT'_3$. We commence with the calculation of $vb' = {M'}_{i_b} \cdot \overrightarrow{u'} \rightarrow CT'_2 = \{B'_b, C'_b\}$:
  + Define $\rho'(i'_b) = b$ so that it aligns with the new policy. Randomly choose $r'_{i_b} \in Z^*_p$.
  + Compute $B'_b = g^{vb'} \cdot H(b)^{-r'_{i_b}}$ and $C'_b = g^{r'_{i_b}}$.
- Return the re-encrypted ciphertext to DR: $CT' = \{CT'_1, CT'_2, CT'_3\}$.

### F. *Decrypt($PK_A$, $SK_A$, CT/CT') → $K_{sym}$*

We apply Lagrange interpolation to decrypt leaf nodes and non-leaf nodes. Then, we use those intermediate results to successfully recover the symmetric key $K_{sym}$.

- Download the ciphertext:
  + <u>DR in the same domain with DO</u>: Simply download and decrypt CT, which is the original ciphertext. Decryption occurs if and only if the set of attribute $A$ **satisfies** DO's access policy $(M, \rho)$.
  + <u>DR in a different domain with DO</u>: Download the re-encrypted ciphertext CT from the proxy. Decryption occurs iff the set of attribute $A'$ **satisfies** DR's access policy $(M', \rho')$.
- Decryption:
  + Each leaf node receives a share $v_i$. Its polynomial is defined as: $q_i(x) = k \cdot q'_i(x) \rightarrow q_i(0) = k \cdot q'_i(0) = k \cdot v_i$ where $q'_i(x)$ is the original share-generating polynomial. In this context, $q_i(0)$ means the **user-specific exponent** of the global secret share.
  + Define a decryption function for leaf nodes: $F_i = \frac{e(SK_{3,i}[1], B_i)}{e(SK_{3,i}[0], C_i)} = \frac{e(g^k \cdot {W_i}^{k_i}, g^{v_i})}{e(g^{k_i}, {W_i}^{v_i})} = \frac{e\left(g^k \cdot (g^{p_i})^{k_i}, g^{v_i}\right)}{e\left(g^{k_i}, (g^{p_i})^{v_i}\right)}. = \frac{e(g,g)^{v_i(k+p_i k_i)}}{e(g,g)^{k_i p_i v_i}} = e(g,g)^{k \cdot v_i} = e(g,g)^{q_i(0)}$
  + For non-leaf nodes $w$, let the set of children of w is S = {index(z) : z ∈ S ∩ z is satisfied}. Starting from $F_z$, we have to map each of its leaf nodes in order to reconstruct a group of elements of the form $e(g, g)^{kv}$ using Lagrange interpolation. Its bilinear form can be represented as:

$$F_w = \prod_{i \in S} {F_i}^{\Delta_{i,S}(0)} = \prod_{i \in S} [e(g,g)^{k \cdot v_i}]^{\Delta_{i,S}(0)}$$

$$= \left[e(g,g)^{k \sum \ v_i}\right]^{\Delta_{i,S}(0)} = e(g,g)^{k \sum \ v_i \cdot \Delta_{i,S}(0)}$$

By definition of Lagrange interpolation, the new secret key to be share $s'$ can be reconstructed using: $s' = \sum \ v_i \cdot \Delta_{i,S}(0)$ where:

- $v_i = q(x_i)$, a value evaluated at a point from the polynomial $q(x)$.
- $\Delta_{i,S}(0) = \prod_{j \in S \neq i} \frac{0 - x_j}{x_i - x_j}$ is the Lagrange basis coefficient.

Eventually, we have $F_w = e(g,g)^{k \cdot s'}$, which is in the desired form $e(g,g)^{k \cdot v}$.

- Recover the symmetric key:

DR uses $F_w$ and the input to reconstruct the symmetric key. Recall from our input:

+ $CT_1 = \{A_1, A_2\} = \{K \cdot e(g,g)^{mv}, g^{nv}\}$;
+ $SK_1 = g^{\frac{m+k}{n}}$.

To recover the symmetric key, perform an operation:

re-encrypt(CT= $\{K \cdot e(g,g)^{mv}, g^{nv}\}$) = $K \cdot e(A_2, SK_1)$=

$$\frac{A_1 \cdot F_W}{e(A_2, SK_1)} = \frac{K \cdot e(g,g)^{mv} \cdot e(g,g)^{k \cdot v}}{e\left(g^{nv}, g^{\frac{m+k}{n}}\right)} = K \frac{e(g,g)^{mv+kv}}{e(g,g)^{v(m+k)}} = K$$

## 3.3 Security Proof and Analysis

### 3.3.1. IND-CPA secure in BDHE assumption

We aim to establish IND-CPA security [5] for our scheme, particularly, developing *Indistinguishability under Chosen-Plaintext Attack*. In other words, our scheme is IND-CPA secure if an attacker cannot figure out which message was encrypted, even though they know beforehand:

- The public key *pk* of the scheme.
- The message to be encrypted of their choice.
- A challenge ciphertext $CT'$ to encrypt either $M_0$ or $M_1$.

We further enhanced the security of our scheme with the *Bilinear* Diffie-Hellman Exponent (BDHE) assumption. It states that:

"*If the decisional BDHE problem is hard in bilinear groups (G, $G^T$), there is no probabilistic polynomial-time (PPT) adversary A that can distinguish between the encryption of any two chosen messages of equal length*."

By integrating BDHE, we can prove the IND-CPA security of our scheme. That means, if an attacker could break IND-CPA, they could also solve the BDHE assumption, which is assumed to be hard. It would lead to a contradiction, our scheme is therefore to be secured under BDHE assumption.

***Proposition 3.3.1.1*** **(IND-CPA security achievement)** If the decisional BDHE problem is hard in bilinear groups ($G$, $G^T$), there is no probabilistic polynomial-time (PPT) adversary $A$ that can distinguish between the encryption of any two chosen messages of equal length.

### 3.3.2. The IND-CPA game

To provide a sketch of proof for **Proposition 3.3.1.1**, we introduce the security game of the applied IND-CPA:

**A. *The BDHE challenge***

The goal of the adversary $A$ (the attacker) is to distinguish between the values of the term $T$ whether it is a valid pairing $e(g,g)^{a^{l+1}}$ or a random element in $G^T$. The challenger $C$, which is used to simulate the environment, needs to embed the BDHE challenge instance into the system's public key and ciphertext so that:

+ Everything looks legitimate to $A$. That means, we need to reconstruct the challenge instance that shares a common form of our entities in the scheme.
+ The challenge ciphertext hides one of two messages under the BDHE term which is the tree representative access policy denoted as

$$T = e(g,g)^{a^{l+1}} \quad (1)$$

**B. *Simulation setup***

The challenger generates keys and sends the public keys *pk* to the adversary. The master secret key *sk* is unknown to $C$.

**G. *Query Phase 1***

If $A$ satisfies the challenge policy ($M^*$, $\rho^*$), $C$ randomly selects an output from {0,1} and ends the query. Otherwise, the challenger simulates the key (performs KeyGen($A$)) using random values followed by the simulation of ReKeyGen without the secret key *sk*. Note that the simulation just needs to be indistinguishable from a valid key.

**C. *Challenge Phase***

The adversary $A$ submits two equal-length messages $M_0$ and $M_1$. The challenger then chooses a bit $b \in \{0,1\}$ and encrypts the corresponding message $M_b$. It then constructs the challenge ciphertext $CT' = \{CT'_1, CT'_2, CT'_3\}$. The encrypted message is hidden in $CT'_1 = \{A'_1, A'_2\} = \{M_b \cdot T, g^{nv}\}$. In the real encryption phase, $A_1 = K \cdot e(g,g)^{mv}$, and $A'_1$ acts as the masking in the challenge simulation. If the adversary can distinguish between $M_0 \cdot T$ and $M_1 \cdot T$, it will contradict our BDHE assumption. Finally, $C$ outputs the challenge ciphertext to $A$.

**D. *Query Phase 2***

The adversary $A$ may continue to ask for keys of other attributes if they still do not satisfy the challenge policy ($M^*$, $\rho^*$), or ask for re-encrypted keys. It then performs message encryption similar to Query Phase 1 for other messages (not $M_0$ and $M_1$).

**E. *Guess and Prediction***

The adversary $A$ outputs a guess $b' \in \{0,1\}$ of $b$. If $b' = b$, the adversary "wins" and $C$ outputs 0, which means $T$ is a valid pairing: $T = e(g,g)^{a^{l+1}}$. Otherwise, $C$ outputs 1 indicating $T$ is a randomized factor. If the adversary cannot guess $b$ with the probability significantly better than 50%, the scheme is IND-CPA secure. Mathematically, the advantage of the adversary $A$ can be written as:

$$\mathrm{Adv}_A = \left| Pr[b' = b] - \frac{1}{2} \right|$$

# 4. Numerical results and discussion

## 4.1 Testbed setup

We utilized the package `py_ecc` [6] to perform elliptic curve operations and bilinear pairing $e(g, g)$ simulation. We encrypt the input message, then we get the mean execution time of the entire scheme (in *ms*) with the average winning rate of the scheme using the following sets of attributes:

$A_1$ = ["Doctor", "Professor", "Researcher"]

$A_2$ = ["Doctor", "Student", "Professor", "Researcher"]

In the first half of the table, we encrypt the message *message* = "OISP Symposium". In the other half, we encrypt a longer message *message* = "The 9th Student Conference". Each time, we take 10 samples and measure the average execution time (rounded to 4 decimal places) and the mean winning rate of the scheme.

## 4.2 Computation time and win rates

Table 1 and 2 summarize the performance of the two scenarios deployment performance.

| Number of trials | $A_1$ | $A_2$ | $A_1$ | $A_2$ |
|---|---|---|---|---|
| 1000 | 415.6945<br>49.82% | 479.7816<br>49.96% | 372.6626<br>50.56% | 487.7665<br>49.78% |
| 1500 | 368.3322<br>49.99% | 444.7227<br>49.52% | 369.3436<br>50.06% | 479.6716<br>49.69% |
| 2000 | 383.3293<br>50.55% | 450.3697<br>50.00% | 421.1391<br>50.17% | 486.6117<br>49.84% |

*Table 1: Results from Google Colab*

| Number of trials | $A_1$ | $A_2$ | $A_1$ | $A_2$ |
|---|---|---|---|---|
| 1000 | 453.6785<br>48.93% | 530.2365<br>50.32% | 434.6213<br>49.68% | 542.0163<br>49.58% |
| 1500 | 431.0967<br>49.75% | 547.3329<br>49.67% | 459.9322<br>48.99% | 553.5672<br>49.96% |
| 2000 | 448.5875<br>49.99% | 531.3054<br>49.92% | 444.9861<br>49.65% | 554.5456<br>49.37% |

*Table 2: Results from local machine*

### 4.2.1. Platform comparison

We spotted a significant distinction of the execution time between Google Colab and our virtual environment on the local machine. Furthermore, a slight difference in the win rate of the adversary interests us. We therefore construct the following table to dive in further comparisons of the two platforms:

| Observation | Google Colab | Local Machine |
|---|---|---|
| Lower execution time | ✓ | ✗ |
| Win rate ≈ 50% | ✓ | ✗ |
| Consistent execution time across trials | Generally ✓ | More variability |

Google Colab is a cloud-based platform that uses highly optimized, often GPU-backed, virtualized environments with efficient random number generation and integer arithmetic. Meanwhile, the local machine probably has less efficient Python math libraries for big integer arithmetic, possible CPU resource constraints and higher overhead in cryptographic simulations (hashing, elliptic curve operations, for instance).

### 4.2.2. Message and access control impact

We would like to see the impact of different lengths of the input message on the execution time of the scheme and the accuracy of the adversary. From the above numerical results, we conclude that:

| Message | Execution Time | Win Rate |
|---|---|---|
| "OISP Symposium" | ✓ | ✗ |
| "The 9th Student Conference" | ✗ | ✓ |

While longer messages entail higher modular operations and costlier pairing operations, the execution time therefore increases. However, the accuracy of the scheme likely depends on the randomness and sampling of *b*' rather than the length of the chosen message. But in general, longer messages slightly improve the noise in the scheme, potentially improving distinguishability in IND-CPA tests.

### 4.2.3. The size of attribute set

To evaluate the impact of attribute set size on the execution time, we conduct a workload of larger attribute set to determine whether there is a tremendous difference between the run time of the scheme and its winning rate:

| Attribute Set | Set Size | Execution Time | Win Rate |
|---|---|---|---|
| $A_1$ | 3 | ✓ | ✓ |
| $A_2$ | 4 | ✗ | ✗ |

It is obvious that larger size of the set will lead to an increase in the number of keys to be generated, thus the number of elliptic curve operations goes up. The computation cost also rises since each attribute generates a random secret $W_i$ and a combined element $g^{r_1} \cdot W_i^{r_i}$. In conclusion, the larger size of the attribute set will diminish the execution time of the scheme.

On the other hand, the security of the scheme slightly improves. The drop in the win rate indicates an increase in key generation complexity, causing more noises in the simulation of the adversary.

## 5. Conclusions

Our proposed work has established a comprehensive cross-domain access control scheme. By integrating a practical Odoo-based access control mechanism and proposing a secure PRE-based scheme for cross-domain attribute mapping, we gain desired numerical results that demonstrate success in both aspects. The following conclusions can be drawn from these results:

+ The Odoo-based inter-domain access control mechanism has been effectively integrated into the ERP environment. It enables subcontractor collaboration across independent sites, ensuring that tasks, authentication, and issue tracking can be effectively coordinated. The modular system also supports dynamic delegation and adaptive workflows, therefore makes it applicable to large-scale Construction 4.0 projects.

+ The proposed PRE scheme offers a mathematically and cryptographically secure framework for the attribute-based access control across domains. Under BDHE assumption, the scheme achieves IND-CPA security with different criteria, ensuring the reliability of the encrypted data through challenge simulation. The scheme further supports efficient key generation, attribute mapping, and re-encryption with the help of Lagrange interpolation to achieve better confidentiality.

+ By utilizing the `py_ecc` package for our elliptic curve simulation, we can confirm that:

- The execution time of the scheme reasonably scales with the increasing size of the attribute set and message length.
- The winning rate of the scheme remains consistent and ideal, which approximates 50%. As a result, it helps validate our proof and improve IND-CPA security across domains.
- Cloud-based execution (for instance, Google Colab) outperforms local setups to some extent. This also implies the feasibility for deployment in distributed cloud environments.

**Acknowledgement:** This research is funded by MURATA SCIENCE FOUNDATION and Ho Chi Minh City University of Technology (HCMUT), VNUHCM under grant number **24VH05**. We acknowledge MURATA and Ho Chi Minh City University of Technology (HCMUT), VNUHCM for supporting this study.